\documentclass[a4]{article}

\usepackage[margin=3cm]{geometry}

\usepackage[none]{hyphenat}

\usepackage{url}

\usepackage{multirow}
\usepackage{epsfig}
\usepackage{subfigure}
\usepackage{color,pstricks,graphicx}
\usepackage[normalem]{ulem}

\begin{document}

\title{Finding representative sets of optimizations for adaptive multiversioning
applications\thanks{This work was
supported by a grant from the National Natural Science Foundation of
 China (No.60873057). This work was also partially supported by the MILEPOST project~\cite{milepost}.}}

\author{
Lianjie Luo$^{1,2}$, Yang Chen$^{1,2}$, Chengyong Wu$^1$, Shun Long$^3$, Grigori Fursin$^4$ \\
\\
  $^1$~Key Laboratory of Computer System and Architecture, \\
 Institute of Computing Technology, Chinese Academy of Sciences, Beijing, China \\
  $^2$~Graduate School of the Chinese Academy of Sciences, Beijing, China \\
  $^3$~Department of Computer Science, JiNan University, Guangzhou, China \\
  $^4$~INRIA Saclay, Orsay, France~~~\textit{(contact: grigori.fursin@inria.fr)} \\
}

\maketitle

\centerline{\textbf{\textit{
3rd Workshop on Statistical and Machine Learning Approaches}}}
\centerline{\textbf{\textit{
Applied to Architectures and Compilation (SMART'09),}}}
\centerline{\textbf{\textit{
co-located with HiPEAC'09 conference, Paphos, Cyprus, 2009}}}


\begin{abstract}
Iterative compilation is a widely adopted technique to optimize
programs for different constraints such as performance, code size
and power consumption in rapidly evolving hardware and software
environments. However, in case of statically compiled programs, it
is often restricted to optimizations for a specific dataset and may
not be applicable to applications that exhibit different run-time
behavior across program phases, multiple datasets or when executed
in heterogeneous, reconfigurable and virtual environments. Several
frameworks have been recently introduced to tackle these problems
and enable run-time optimization and adaptation for statically
compiled programs based on static function multiversioning and
monitoring of online program behavior. In this article, we present a
novel technique to select a minimal set of representative
optimization variants (function versions) for such frameworks while
avoiding performance loss across available datasets and code-size
explosion. We developed a novel mapping mechanism using popular
decision tree or rule induction based machine learning techniques to
rapidly select best code versions at run-time based on dataset
features and minimize selection overhead. These techniques enable
creation of self-tuning static binaries or libraries 
adaptable to changing behavior and environments at run-time
using staged compilation that do not require complex recompilation frameworks 
while effectively outperforming traditional single-version non-adaptable code.

\end{abstract}


\section{Introduction}

The past two decades have seen a rapid evolution of architectural
designs and growth of their complexity. Modern compilers employ many
advanced optimizations to achieve better performance on such
architectures. However, they often fail due to simplified hardware
models used for static analysis and a lack of run-time information.
Iterative compilation became a widely adopted technique to optimize
programs for different constraints such as performance and code size
without a priori knowledge of the underlying
hardware~\cite{CSS99,pfdc,FOK02,TVVA03,cooper2005a,vista,PE2006,HB2006,HE2008,esto}.
However, it is often used to optimize programs for a specific
dataset which may not be practical as shown in~\cite{FCOP2007} where
an influence of multiple datasets on iterative compilation has been
studied using a number of programs from MiBench benchmark suite.

Hybrid static/dynamic optimization approaches have been introduced
to tackle those problems and allow compilers make better
optimization decisions at run-time. Search-based methods have been
adopted in several well-known library generators such as
ATLAS~\cite{atlas}, FFTW~\cite{fftw} and SPIRAL~\cite{spiral} to
identify different optimization variants for different inputs that
fit the computer architecture best at run-time. Some more general
approaches have also been introduced
in~\cite{Byler1987,Diniz1997,Voss2001,LAHP2006} to make static
programs adaptable at run-time by generating different code versions
statically or dynamically and selecting them based on a given
context, performance prediction or according to the changing
run-time behavior. However, most of these frameworks are limited to
only a few optimizations and do not have mechanisms to select a
representative set of optimization variants. ~\cite{FCOP2005}
presents a framework which creates adaptive binaries and statically
enables run-time adaptation based on function multiversioning,
iterative compilation and low-overhead hardware counters monitoring
routines. It searches for complex combinations of optimizations in
an off-line iterative manner. However, it is based on a reactive
model and provides no pruning mechanism in order to avoid code size
explosion.

This paper presents a novel approach to generate only a limited
number of representative optimization variants across all datasets
without performance loss or code-size explosion. It is based on
finding good optimizations for hot program or library functions with
different datasets using traditional off-line random iterative
compilation in large optimization spaces and then iteratively
pruning those variants while controlling overall performance and
code size. When representative set of optimizations is found, we
utilize several standard classification algorithms (decision trees
or rule induction) to correlate some characteristics of the datasets
with the best optimized function version. The learned decision trees
or rules are then converted into executable code for runtime version
selection. We evaluated our techniques using Open64 research
compiler and plan to implement this framework inside GCC. However,
hand-written optimization versions, libraries or versions generated
using other optimization techniques or even compiled for different ISA 
(in virtual or heterogeneous environments) can be easily plugged into our framework.

The paper is organized as follows. Section~\ref{sec:framework}
provides a motivation example for multiversioning and outlines the
proposed framework. Section~\ref{sec:representative} describes a
heuristic to find a representative set of optimization versions while
maximizing performance and minimizing code size.
Section~\ref{sec:mapping} evaluates different machine learning
models to map some dataset characteristics to the selected versions
to maximize overall performance and minimize overheads.
Section~\ref{sec:related} summarizes related work in this area,
before concluding remarks and future work in Section~\ref{sec:conclusions}.


\section{Static Multiversioning Framework to Enable Run-time Adaptation}
\label{sec:framework}

\subsection{Motivation}

\begin{figure*}[htb]
  \centering
  \includegraphics[width=6in]{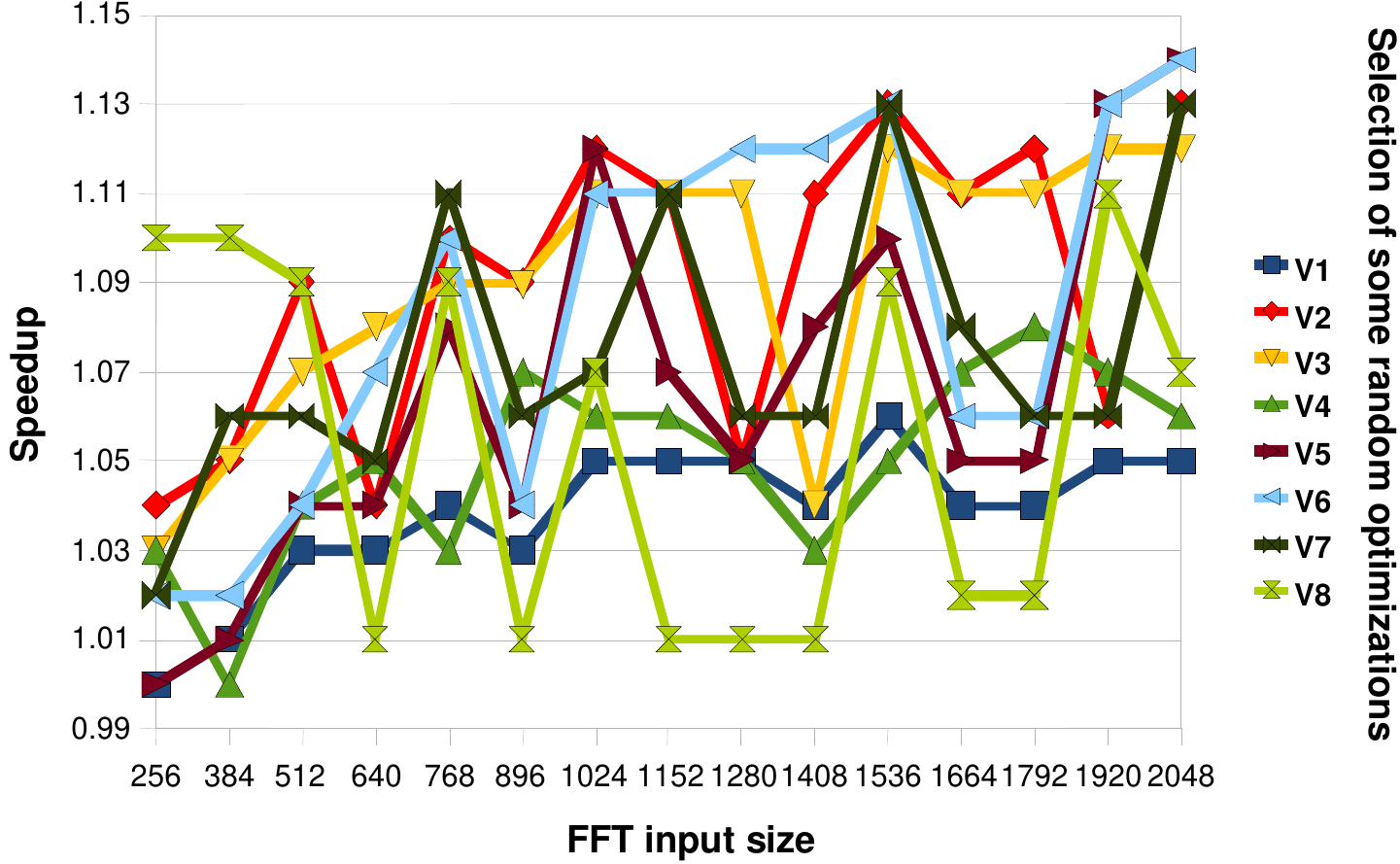}
  \caption{\it Speedup variation of 8 different optimization variants of FFT kernel
for different input sizes on  Opteron 2.6GHz machine using Open64
compiler V4.0 with ICI and with -O3 default optimization level}
  \label{fig:motivation}
\end{figure*}

Some prior works show that different optimization combinations are
needed for kernels or programs with multiple
datasets~\cite{atlas,fftw,FCOP2007}. We decided to confirm these
findings using FFT benchmark with 15 different input sizes on a
recent architecture such as dual-core AMD Opteron 2.6GHz with RedHat
Linux AS4. We use Open64 4.0 compiler with the Interactive
Compilation Interface~\cite{open64-ici} to apply combinations of
fine-grain transformations such as loop tiling and unrolling with
random parameters to the most time consuming loops in the kernel. We
found 8 best optimization variants across those datasets.
Figure~\ref{fig:motivation} shows the speedups over -O3 optimization
level of Open64 for these optimization variants, and how they vary
across different datasets. It shows that the relative performance of
a version can vary significantly on different input data, and no one
single version can outperform all the other versions across all
datasets. This motivates us to develop an automatic adaptive
multiversioning technique which can select a proper version based on a
given runtime context. Moreover, having all 8 versions may not be
practical due to considerable code-size increase and hence pruning
technique is needed to select a small representative set of such
optimizations, when given thresholds for tolerable code size
increase and performance loss across all datasets.

\subsection{Framework}

The aim of our work can be formulated as a multi-objective problem
as follows: given a set of semantically equivalent but differently
optimized versions of a given program, kernel or library function
(or compiled for different architecture in heterogeneous and virtual
environments) for multiple input datasets, find the smallest subset
of versions while maximizing overall performance (or reduce power
consumption for example) and minimizing code size. When such set is
found, use machine learning to build the mapping between dataset
features (or run-time context such as hardware counters) and all
versions in this representative set while minimizing the decision
tree and hence run-time overhead using machine learning.

\begin{figure*}[htb]
  \centering
  \includegraphics[width=3.8in]{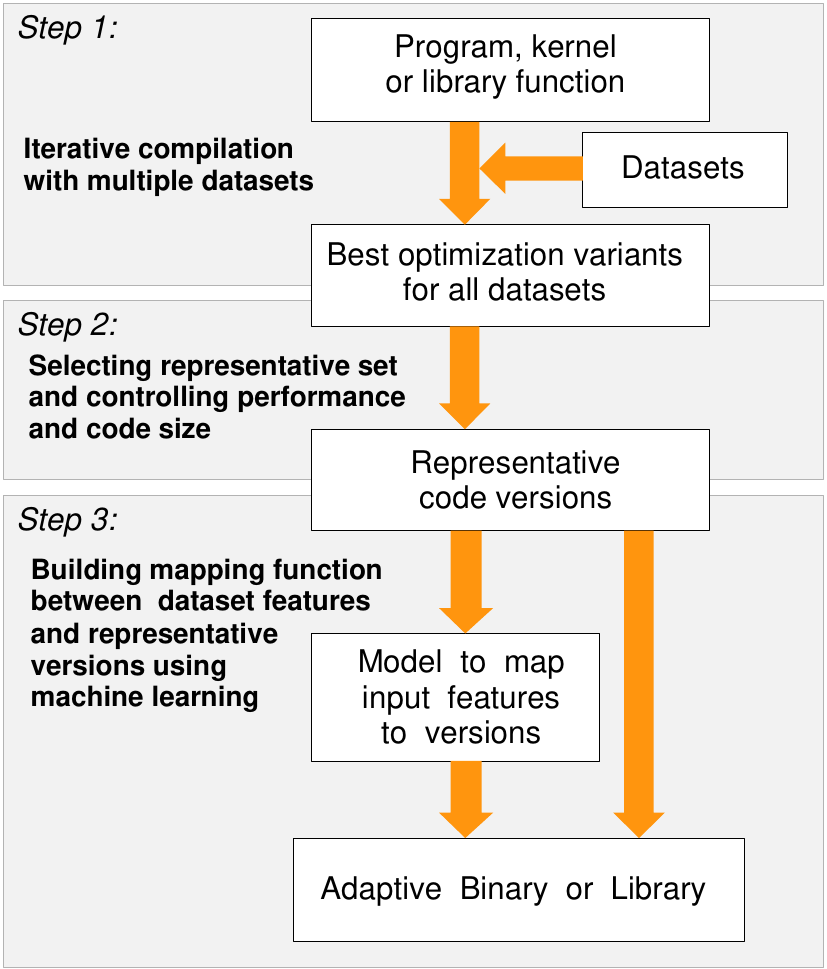}
  \caption{\it Static multiversioning framework to enable run-time optimization and adaptation
while pruning the number of versions, monitoring overall performance, reducing code size
and finding mapping between dataset characteristics and representative versions}
  \label{fig:framework}
\end{figure*}

The framework implementing this approach is presented in
Figure~\ref{fig:framework}. It takes three steps to achieve these
goals. First, we evaluate a large number of combinations of
optimizations for a given program with multiple datasets using
traditional iterative compilation techniques. This step can be
considerably accelerated using techniques such as collective
optimization~\cite{FT2009,cccpf}. Then, we use a heuristic presented
in Section~\ref{sec:representative} to select the representative
versions. Finally, we build a model to map features of a program
input to the representative versions using traditional machine
learning techniques as described in Section~\ref{sec:mapping}. All
representative versions with the resulted selection mechanism are
statically linked into the final executable or library.


\subsection{Adaptive Binaries and Libraries}

\begin{figure*}[htb]
  \centering
  \includegraphics[width=6in]{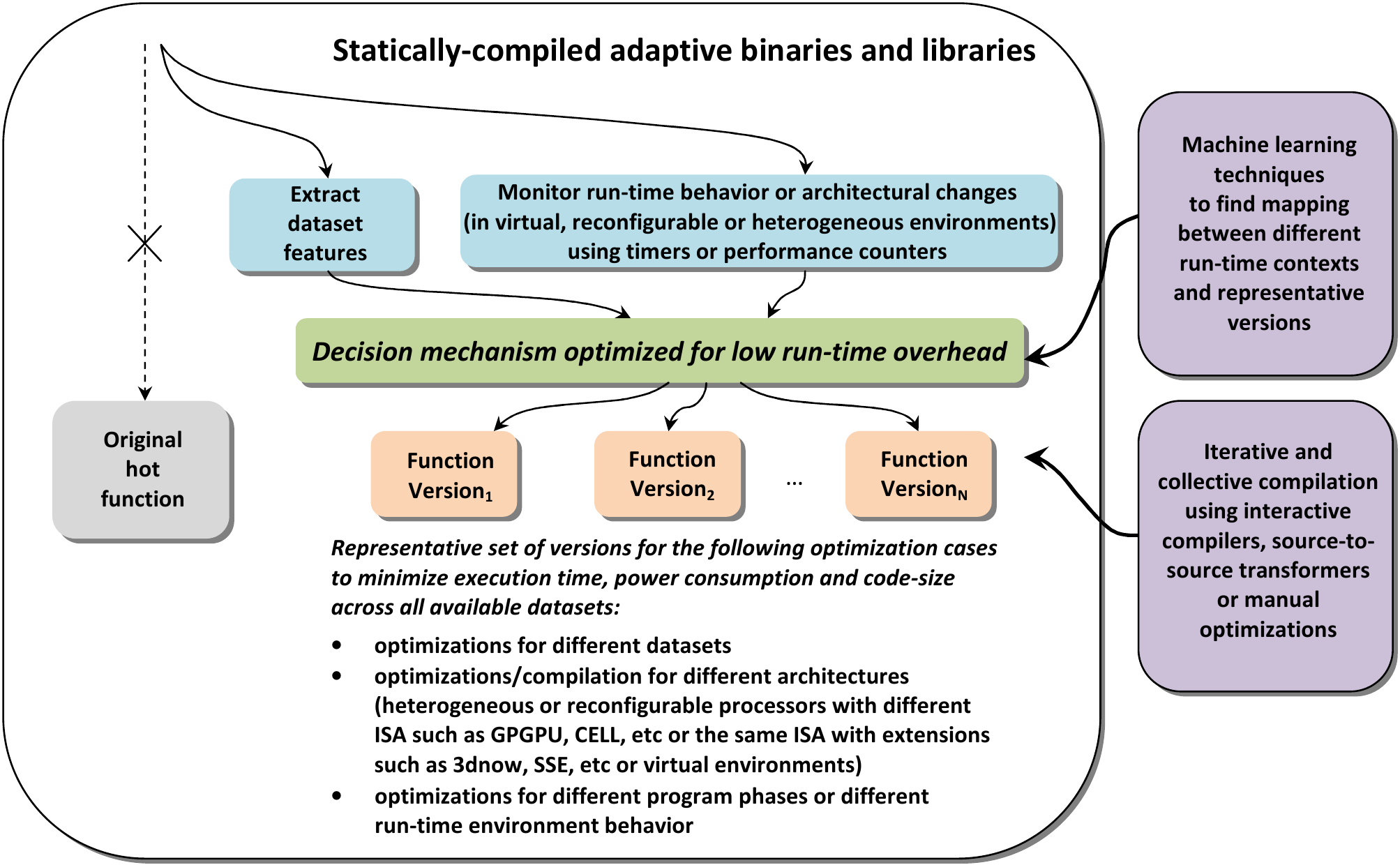}
  \caption{\it Adaptive binaries or libraries with a representative set of multiple function versions
optimized or compiled for different run-time cases and with the
decision tree or rule induction to map them with different run-time
contexts}
  \label{fig:adaptive_program}
\end{figure*}

When the representative set of versions for different run-time
optimization cases is selected and the mapping function is prepared,
we produce the final adaptive binary or library as shown in
Figure~\ref{fig:adaptive_program}. Such binaries or libraries
include run-time routines for dataset/program/environment feature
extraction and programs runtime behavior monitoring in order to
select appropriate versions to improve performance, reduce power
consumption or improve reliability, etc. Though compiled statically,
this code is now adaptable to different datasets, run-time program
and system behavior or even different heterogeneous, reconfigurable
and virtual architectures.

\section{Selection of Representative Optimization Versions}
\label{sec:representative}

\subsection{Experimental Setup and Iterative Compilation}
\label{sec:exp setup}

Iterative compilation is traditionally performed using global
compiler flags or source-to-source transformation tools which is not
always satisfactory, particularly for function-level optimizations.
Interactive Compilation Interface (ICI) has been recently
introduced~\cite{FMTP2008,open64-ici,gcc-ici} to enable fine-grain
optimizations in production compilers with the ability to select
different combinations, phase orders and parameters of available
transformations. We decided to use Open64 4.0 compiler with ICI
enabled~\cite{open64-ici} since it is a well-known research compiler
with multiple aggressive optimizations available. We evaluated the
following transformations using hill-climbing search similar
to~\cite{FOK02}.

\begin{itemize}
\item loop tiling (2..512)
\item register tiling (2..8)
\item loop unrolling (2..16)
\item loop vectorization
\item loop interchange
\item loop fusion
\item array prefetching (8..128)
\end{itemize}

To validate our framework, we selected two widely used kernels:
DGEMM from level-3 BLAS~\cite{BLAS} library of NetLib, and FFT from
UTDSP~\cite{UTDSP}. We evaluated 100 different combinations of
optimizations on DGEMM and 280 on FFT (this number is program
dependent). We randomly generated 1000 distinct datasets for BLAS
and 280 for FFT with different input sizes and data values. All
experiments are performed on a dual-core AMD Opteron at 2.6GHz, with
64KB L1 cache and 1MB L2 cache for each core, and 16GB memory,
running RedHat AS4 (kernel 2.6.9-42).

\subsection{Heuristic to Select the Representative Set of Optimizations}

\begin{figure*}[htb]
  \centering
  \includegraphics[width=4.6in]{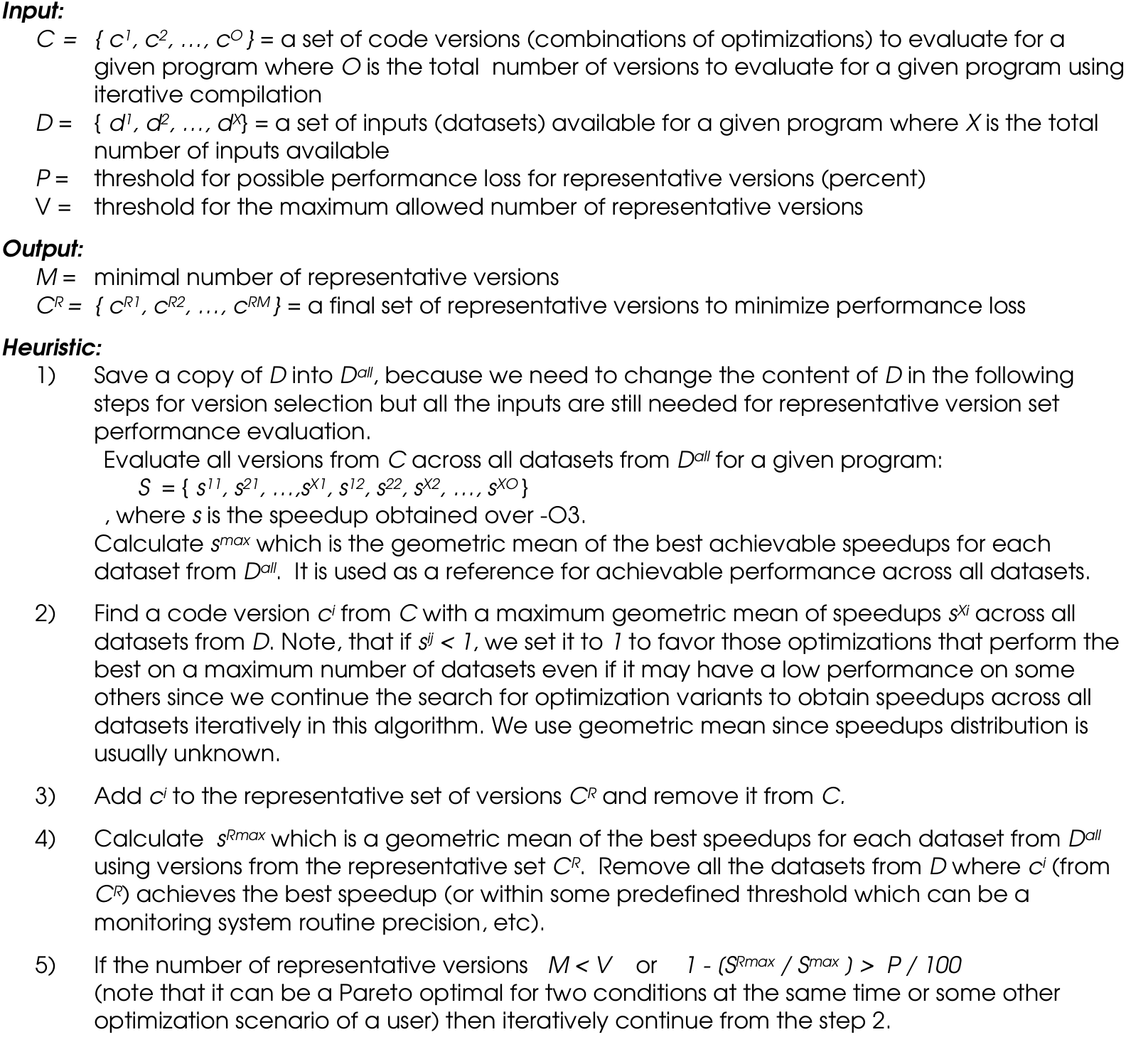}
  \caption{\it Algorithm to find a minimal representative set of versions
that minimizes performance loss across all datasets}
  \label{fig:adaptive_algorithm}
\end{figure*}

Given a potentially large number of combinations of optimizations
(code versions) for a set of sample inputs, we would like to select
only a minimal set of representative ones that obtain best
performance on a maximum set of inputs. A heuristic algorithm for
this is presented in Figure~\ref{fig:adaptive_algorithm}. Depending
on the user's optimization priority (overall achieved performance vs
maximum number of allowed versions to control code-size explosion),
this algorithm tries to prune these versions and leaves only the
representative ones that outperform the single version code for as
many datasets as possible. To achieve this, it adopts a greedy
strategy that is explained in Figure~\ref{fig:adaptive_algorithm}. 
And we plan to use multi-objective Pareto optimization
for optimal performance/code size ratio as described
in~\cite{HB2006,HE2008} in the future work. Further more, we plan to
extend the presented heuristic to take into account the mapping
mechanism (described later in this paper) to be sure that all the
representative versions can be effectively mapped with the dataset
characteristics.

Using our algorithm in Figure ~\ref{fig:adaptive_algorithm} we
obtained 3 representative optimization versions for DGEMM and 4 for
FFT. Evaluation of overall obtained performance and different
overheads for these programs is presented in Sections~\ref{sec:perf}
and~\ref{sec:overhead} respectively.


\section{Run-time Version Mapping Mechanism}
\label{sec:mapping}

\subsection{Objective}

To be able to statically create adaptive applications and libraries,
we need an effective mechanism to select the representative
optimization versions at run-time based on dataset characteristics
or other run-time context such as dynamic feature vector of
performance counters, information about process migration in
multi-core heterogeneous architectures and virtual environments.
Machine learning~\cite{Mitchel1997} has been effectively used to
learn and build such mappings automatically. We evaluated some
commonly used classification algorithms available in the popular
WEKA~\cite{WEKA} machine learning suite that supporting multiple
standard techniques such as clustering, classification, and
regression.

All these algorithms vary in applicability and complexity depending
on the problem encountered. In order to find one suitable for our
version mapping mechanism, we decided to evaluate two widely adopted
methods with several variants: direct classification (DC) and
performance prediction model (PPM). Given a test case of a dataset,
DC returns the most similar case from its prior experience (the
training set), i.e. the optimization for another dataset most
similar to the given one, expecting that the speedup will also be
similar. On the contrary, PPM usually uses a probabilistic approach
to correlate dataset features with available optimizations and
speedups, uses probability distribution to suggest a set of good
optimizations for a dataset before selecting the best out of these
optimizations. Generally speaking, PPM performs better than DC, but
at the extra cost in performance estimation, prediction and
comparison. In our case, though the training cost can be tolerated ,
it is critical to link an optimized run-time decision tree to the
adaptive binary or library in order to select appropriate versions
online without considerable overhead. Six most commonly used
heuristics for DC and PPM are evaluated, among which the best is
selected.

It is vitally important for all machine learning techniques to find
the suitable characterization of datasets in order to correlate
dataset attributes with influential optimizations. It is a
challenging task and beyond the scope of this paper. As a first
step, we decided to use only the dimensions of input arrays since
they are known to influence most of the transformations evaluated in
this paper. There are other attributes that we plan to use in the
future, such as the values of the entire array. However, it
inevitably leads to a larger number of attributes to consider and
may result in overfitting, while Li~et~al~\cite{Li2004} suggested
that the characteristics of input array elements may not be as
important as the distribution of the values of them. Implicit
attributes such as pointer type could also describe programs and
library functions. However, if they point to an array, it may not be
enough to learn from the value of the pointer itself. Therefore, we
should consider high-level information about loops and array
dimensions. We plan to combine all these characteristics with
dynamic attributes such as performance counters, available hardware
and software resources, system workload in the extension of this
work. We can find or even generate as many features as possible and
then automatically find the important ones using standard machine
learning techniques such as Principle Component Analysis in order to
keep the number of attributes low while maintaining the accuracy of
learning and prediction.

\subsection{Evaluation of Direct Classification vs Performance Prediction Model}

Six different learning methods are adopted in DC~\cite{Witten2005,WEKA}:

\begin{itemize}
\item SMO - Support Vector Machine based
\item J48 - decision tree based
\item REPTree - decision tree based
\item JRip - rule based
\item PART - rule based
\item Ridor - rule based
\end{itemize}

\begin{figure*}[htb]
\centering 
 \includegraphics[width=3.6in]{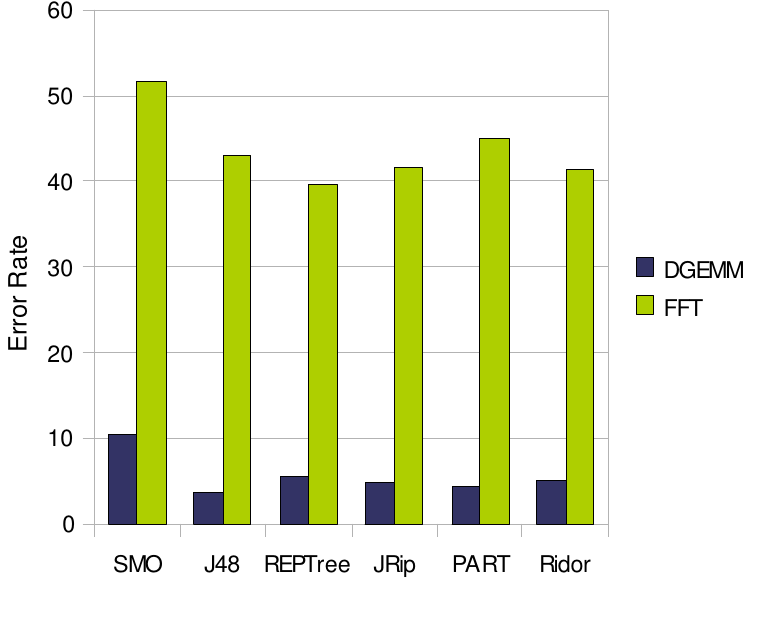}
\caption{Error rate of Direct Classification}
\label{fig:eval_dc}
\end{figure*}

We used a representative set of optimizations, a large number of
datasets in our experiments, which were carried out in a ten-fold
cross validation. We decided to use array dimension as the dataset
characterization as mentioned earlier. Error rate is used as the
performance evaluation metric for DC, and root relative squared
error~\cite{Witten2005} for PPM which are standard metrics for these
algorithms in WEKA.

Figure~\ref{fig:eval_dc} shows that the classification accuracy
depends on the given program and a machine learning method. J48
achieves the lowest error rate for DGEMM, while REPTree minimizes it
for FFT. It is worth noting that the error rate of most of the
classification algorithms have a very high error rate for FFT, more
than 40\% in most cases. This could be caused by a poor dataset
characterization which needs further study in the future.

Six different learning methods are available for PPM:

\begin{itemize}
\item LeastMedSq - linear regression based
\item LinearRegression - linear regression based
\item PaceRegression - linear regression based
\item SMOreg - Support Vector Machine based
\item REPTree - decision tree based
\item M5Rules - rule based
\end{itemize}

\begin{figure*}[htb]
\centering 
 \includegraphics[width=4in]{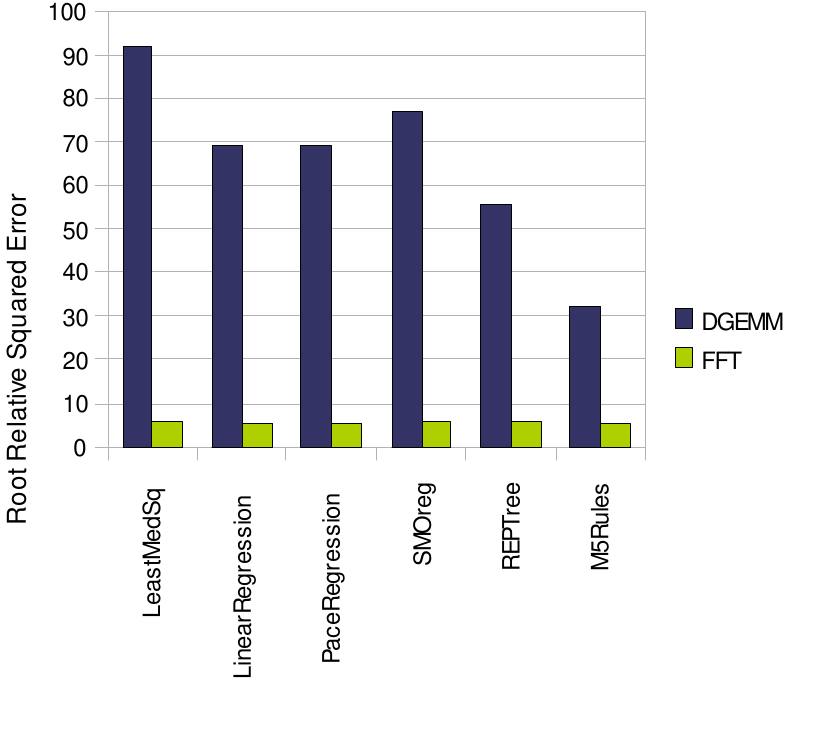}
\caption{Root relative squared error of Performance Prediction Model}
\label{fig:eval_ppm}
\end{figure*}


Figure~\ref{fig:eval_ppm} demonstrates that M5Rules outperforms all
other methods for both DGEMM and FFT.
It is interesting to note that the best performing algorithms from DC and PPM are either
decision tree or rule based which suggests that these methods suit
our mapping objective best. We leave the detailed comparison of
different algorithms for the future work.

%

\subsubsection{Performance Evaluation of the Multiversioning Approach}
\label{sec:perf}

\begin{figure*}[htb]
\centering 
 \includegraphics[width=4.6in]{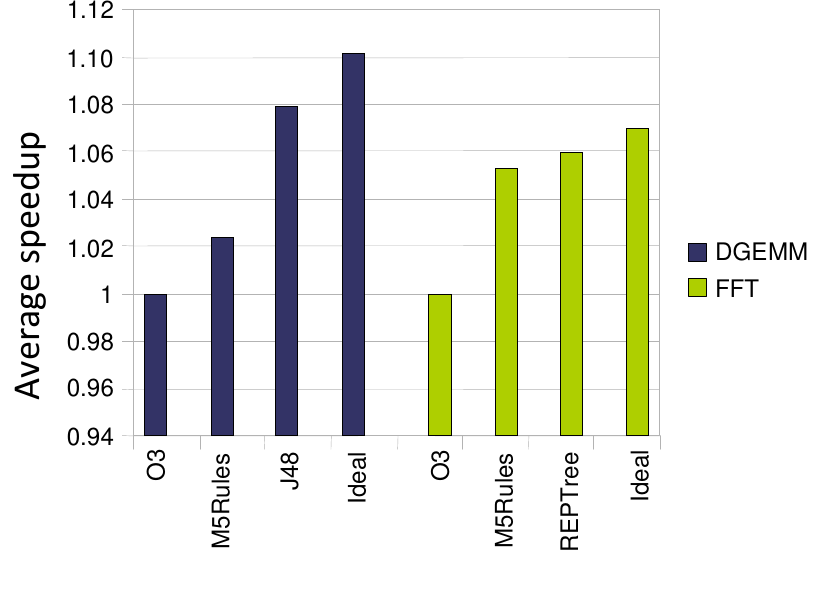}
\caption{Performance of DGEMM and FFT with multiversioning}
\label{fig:performance}
\end{figure*}

Once the best performing mapping algorithm is found, we can evaluate
the mapping in a realistic environment by creating an adaptive
binary or library linked with the selection function and
representative optimization versions (3 versions for DGEMM, 4 for
FFT). Then the produced binaries were executed with randomly
generated test inputs (990 distinct inputs for DGEMM, 82 for FFT,
none of them identical to the training data) on our experiment
platform specified in section~\ref{sec:exp setup}.
Figure~\ref{fig:performance} summaries the performance results which
include the dynamic version selection time (except for the estimated
"ideal" case). The "ideal" numbers in Figure~\ref{fig:performance}
for DGEMM and FFT refers to the estimated average speedup which
could be achieved when the predication accuracy of the machine
learned model for runtime version selection were 100\%. It
demonstrates by combining static multiversioning with dynamic
version selection from 3~4 representative optimizations based on J48
and REPTree mapping mechanism using simple dataset features we can
gain 98\% of the available speedup.

\subsection{Overheads for Code Size and Run-time Selection Mechanism}
\label{sec:overhead}

The introduction of multiple code versions in the binary or library
as well as the run-time version selection inevitably results in code
growth and run-time overhead. Table \ref{tab:overhead} presents the
number of representative versions found for DGEMM and FFT. It
demonstrates that the run-time overhead and the code growth for the
version selection can be negligible whilst the overall code growth
 is not, due to multiversioning. However, depending on
the user optimization scenarios, this overhead could be tolerable or
reduced during the multi-objective tuning of the performance,
representative optimization set and the overall code size. We
believe that such an approach can, without a complex error-prone
dynamic recompilation framework, automatically create static
binaries and libraries which are adaptable to different behavior and
environments at run-time.

\begin{table}[h]
\centering
\begin{tabular}{|p{60pt}|p{60pt}|p{60pt}|p{60pt}|p{60pt}|}
\hline
\parbox{60pt}{\centering

} & \parbox{60pt}{\centering {\scriptsize No of representative versions } } &
\parbox{60pt}{\centering {\scriptsize Run-time selection overhead} } &
\parbox{60pt}{\centering {\scriptsize Code size growth due to selection mechanism} } &
\parbox{60pt}{\centering {\scriptsize Code size growth due to multiversioning}
} \\
\hline
\parbox{33pt}{\centering
{\scriptsize DGEMM} } & \parbox{27pt}{\centering {\scriptsize 3} } &
\parbox{35pt}{\centering {\scriptsize 0.7\%} } &
\parbox{35pt}{\centering {\scriptsize 1.8\%} } &
\parbox{37pt}{\centering {\scriptsize 4.2\%}
} \\
\hline
\parbox{33pt}{\centering
{\scriptsize FFT} } & \parbox{27pt}{\centering {\scriptsize 4} } &
\parbox{35pt}{\centering {\scriptsize 0.5\%} } &
\parbox{35pt}{\centering {\scriptsize 8.8\%} } &
\parbox{37pt}{\centering {\scriptsize 76.5\%}
} \\
\hline
\end{tabular}
\caption{Overheads of the static multiversioning approach}
\label{tab:overhead}
\end{table}

\section{Related Work}
\label{sec:related}

Iterative compilation is an effective technique to optimize programs
on a wide range of different architectures without a priori
knowledge of the hardware/software environment. It is performed in a
feedback directed manner, i.e. the compiler's static optimization
heuristics are replaced with an exploration of an optimization space
, each step of which consists of program compilation, execution
 and search decision revision.

Iterative compilation has been widely used to optimize both kernels
and larger programs on a given
architecture~\cite{pfdc,KKOWb,CSS99,CST02,FOK02,vista,esto,cooper2005a,TVVA03,FOTP2005,FCOP2005,PE2006,HB2006,HE2008,FT2009}
. Various optimization search spaces (composed of various parametric
transformations and of different phase orders) are considered in
order to minimize the execution time or code size. However it is a
very time-consuming process which is unacceptable in many practical
scenarios. To accelerate the iterative search process, most of the
above works use some heuristics to focus optimizations. Machine
learning techniques have been used recently to enable optimization
knowledge reuse, predict good optimizations and speedup iterative
compilation~\cite{Monsifrot,metaOpt03,StephensonA05,CavazosPLDI04,soffa2005,ABCP06,CFAP2007,FMTP2008},
such techniques include genetic programming, supervised learning,
decision trees, predictive modeling etc.

Iterative compilation is usually used to optimize program with one
dataset which is not practical. This is demonstrated
in~\cite{FCOP2007} where the influence of multiple datasets on
iterative compilation has been studied using a number of programs
from MiBench benchmark. Hybrid static/dynamic approaches have been
introduced to tackle such problems. They are used in a well-known
library generators such as ATLAS~\cite{atlas}, FFTW~\cite{fftw} and
SPIRAL~\cite{spiral} to identify different optimization variants for
different inputs to improve overall execution time. Some general
approaches have also been introduced
in~\cite{Byler1987,Diniz1997,Voss2001,LAHP2006} to make static
programs adaptable to changes in run-time behavior by generating
different code versions for different contexts. However, most of
these frameworks are limited to simple optimizations or need complex
run-time recompilation frameworks. None of them provide techniques
to select a representative set of optimization variants.

Another hybrid static/dynamic framework has been introduced
in~\cite{FCOP2005,FT2009} to create self-tuning binaries. Run-time
adaptation is achieved by first using off-line iterative search for
arbitrary combinations of available optimizations and then inserting
into the static binary multiple versions of hot functions as well as
low-overhead hardware counters monitoring routines.

None of the above techniques addresses the issue of automatic
selection of a minimal representative set of optimizations for
kernels or programs with multiple datasets in order to maximize
overall performance and minimize code size explosion. The version
selection mechanisms should be based on program input
characteristics. This paper attempts to address these issues.We
believe that this is an important practical step forward toward
automatic creation of static self-tuning programs or libraries
adaptable to different run-time behavior and environments
automatically and without the help of a complex dynamic
recompilation frameworks.


\section{Conclusions and Future Work}
\label{sec:conclusions}

This paper presented a static multiversioning approach with dynamic
version selection which enables run-time optimizations based on
iterative compilation, dataset characteristics and machine learning.
It is capable of generating static binaries adaptive to different
environments at run-time. We demonstrate that it is possible to
effectively prune a large number of versions optimized for different
datasets in order to build a representative set across available
datasets. This is achieved without considerable performance loss nor
code size explosion, which makes this approach practical. We also
demonstrate how to use popular decision tree and rule induction
classification algorithms to build an effective and low-overhead
run-time mapping mechanism in order to correlate different datasets
and optimized versions from the representative set.

Experimental results on several kernels demonstrate that our
techniques can improve the overall performance of static programs or
libraries with low run-time overhead. We plan to extend our
algorithm to select representative set of optimizations not only
based on performance but also taking into account both dataset
characterization and possible run-time mapping at the same time. We
will investigate the performance of different machine learning
algorithms for run-time version mapping in detail and evaluate them
for different multi-objective optimization scenarios. We plan to
automate dataset and run-time feature generation in order to improve
our version mapping technique. 
We believe that using staged compilation and self-tuning binaries can simplify
automatic adaptation and optimization of the migrated code in virtual heterogeneous
environments. Furthermore, we plan to combine our technique with collective optimization
method~\cite{FT2009,cccpf} and performance counters monitoring routines~\cite{FCOP2005,CFAP2007}
to evaluate it in a large number of heterogeneous, reconfigurable 
and virtual environments.


\section{Acknowledgments}

We would like to thank Yuanjie Huang, Mingjie Xing, Lujie Zhong, and
Liang Peng for their help on the writing of the paper and implementation of the framework.
We would like to thank Olivier Temam and Albert Cohen for fruitful discussions.

\bibliographystyle{abbrv}
\bibliography{fursin}

\end{document}